\documentstyle[12pt,epsfig]{article}
\newcommand{\bbox}[1]{\mbox{\boldmath{$#1$}}}
\def\Vqq{$V_{qq}(\mbox{spin-spin})$}
\def\J2hat{$\hat{\mbox{\boldmath{$J$}}}_{ij}^{(2)}$}
\begin{document}
\today
\begin{center}
\vskip 1.0cm {\Large\bf
Quark-Quark Correlations 
and Baryon Electroweak Observables}
\vskip 1.0cm
{\large H. H\o gaasen\footnote{
E-mail address: hallstein.hogasen@fys.uio.no}, \\
{\large {\it Fysisk Institutt, University of Oslo,
Blindern, 0316 Oslo, Norway}} \\
Kuniharu Kubodera\footnote{
E-mail address : kubodera@sc.edu}, 
and
Fred Myhrer\footnote{
E-mail address : myhrer@sc.edu}
}\\
\vskip 0.8cm
{\large {\it Department of Physics and  Astronomy,
University of South Carolina,\\
Columbia, SC 29208, USA}} \\
\end{center}
\vskip 1.0cm
%


\begin{abstract}
The simple independent quark models have difficulties
explaining simultaneously the totality of the known
hyperon magnetic moments and hyperon
semi-leptonic decay rates.
We show that both the Goldstone boson loop contributions
and two-quark effective exchange currents are essential
in explaining these observables.
\end{abstract}

\vskip 1cm \noindent

Chiral symmetry, one of the basic symmetries of QCD,
plays an important role in hadron physics.
A key point is that chiral symmetry 
is spontaneously broken, resulting in 
the generation of Goldstone bosons,
and these Goldstone bosons have strong influence
on the static properties of hadrons and
low-energy interactions among hadrons.
These effects have been explored with much success 
in, e.g., chiral quark models
and chiral perturbation theory.
The existence of the Goldstone bosons  
leads to a natural picture of a baryon 
consisting of a quark core 
plus a surrounding Goldstone boson cloud, 
with the coupling between the core and the cloud
dictated by chiral symmetry.
It is to be noted that loop diagrams 
involving the Goldstone bosons give rise to
non-analytic corrections to the hadric observables
and that these non-analytic contributions 
are determined model-independently 
from chiral symmetry. 
This feature has recently been exploited 
by Thomas et al. \cite{thomas00} in a calculation 
of the baryon magnetic moments.

Apart from the Goldstone boson loop corrections,
another important effect is expected from QCD;
viz., the effective spin-dependent
quark-quark interaction [denoted by \Vqq\ ] 
arising from gluon exchange is expected 
to affect spin-depedent hadronic 
observables significantly.
Let $\bbox{J}$ represent a current
that describes the response of a baryon
to an external electroweak probe;
$\bbox{J}$ can be an electromagnetic or weak current.
In general, we expect the form,
$\bbox{J}=\sum_{i=1}^3\bbox{J}_i^{(1)}
+\sum_{ij}\bbox{J}_{ij}^{(2)}$,
where $\bbox{J}_i^{(1)}$ is the 1-body current
of the $i$-th quark,
and $\bbox{J}_{ij}^{(2)}$ is the two-body current
involving the $i$-th and $j$-th quarks\cite{foot1}.
In the simplest impulse approximation,
one assumes  
$\bbox{J}=\sum_{i=1}^3\bbox{J}_i^{(1)}$. 
Now, a point to be emphasized here
is that the existence of \Vqq\  automatically 
engenders the corresponding two-body current, 
which we denote by \J2hat.
In most work in the literature,
however, the presence of \J2hat 
has been overlooked \cite{foot2}.
The purpose of this communication is to illustrate  
an important role \J2hat can play 
in describing baryonic electroweak responses.

As a first example that clearly shows the importance 
of the \J2hat contributions, 
we consider what may be called the $\Xi$-$\Lambda$ puzzle,
which is concerned with the magnetic moments
of the octet baryons, $\Xi^-$ and $\Lambda$.
Experimentally, $\mu_\Lambda \simeq -0.613 \pm 0.004$ n.m.\  
and $\mu_{\Xi^-} \simeq -0.6507 \pm 0.0025 $ n.m.,
and hence
$ R\equiv \mu_{\Xi^-} / \mu_\Lambda \simeq$ 
$1.062 \pm 0.005$.
Thus, the experiments clearly indicate
$|\mu_{\Xi^-} | > |\mu_\Lambda|$.
Now, suppose we adopt a naive ``additive" quark model;
namely, we assume that the relevant current 
is given by 
$\bbox{J}=\sum_{i=1}^3\bbox{J}_i^{(1)}$,
and that the three quarks in the baryons
are independent of each other apart 
from trivial spin-flavor factors. 
It can be shown that any additive quark model
inevitably leads to
\begin{equation}
\mu_\Lambda = \mu_s \; \; \; {\rm and } \; \; \;
\mu_{\Xi^-} = \mu_s +\frac{1}{3} \left(\mu_s - \mu_d \right) ,
\label{eq:exp}
\end{equation}
where $\mu_s$ and $\mu_d$ are the magnetic moments 
of the $s$- and $d$-quarks, respectively.
Then, to explain the experimental fact
$|\mu_{\Xi^-}|>|\mu_\Lambda|$,
we must require $| \mu_d|<|\mu_s|$;
recall that $\mu_s$ and $\mu_d$ are both negative numbers.
In the language of the standard 
non-relativistic quark model,
this requirement implies that
the constituent mass of the $d$ quark 
must be larger than that of the $s$ quark,
an obviously untenable conclusion.
Meanwhile, according to 
\cite{hhfm88,kubodera,myhrer83},
the inclusion of Goldstone boson loops
only slightly changes the value of $R$.
Cloet {\it et al.}\cite{cloet02} has 
very recently developed ``chiral phenomenology"
to refine the simplest quark model,
and has reported that,
partly due to the Goldstone-boson cloud,
$R$ can be increased from
$R=0.8$ (the simplest quark-model value)
to $R= 0.99$.
Yet this latest result still leaves unexplained
the experimental fact, 
$|\mu_{\Xi^-}|>|\mu_\Lambda|$.

We now discuss the role of \J2hat 
in the $\Xi$-$\Lambda$ puzzle.
The origin of \J2hat is analogous 
to that of the nuclear exchange current 
due to a nucleon ``Z-graph",
which arises from a relativistic correction
to the nucleon propagator.
It is well known in nuclear physics 
that the nucleon ``Z-graph" 
give significant contributions
to electroweak observables of nuclei.
A ``Z-graph" relevant to \J2hat
involves gluon exchange and quark-antiquark excitation,
as depicted in Fig. 1.
We note that \J2hat contributing 
to the baryon magnetic moment 
is essentially determined by \Vqq\ 
arising from gluon exchange,
and that the strength of \Vqq\ 
can be monitored by the octet-decuplet mass splitting
({\it e.g.}, the $N$-$\Delta$ mass difference).
To show the effects of \J2hat\ 
explicitly, we use here the results 
obtained in the cloudy-bag 
or chiral-bag model.
According to \cite{hhfm88,kubodera},
the inclusion of the \J2hat contribution
changes $\mu_\Lambda$ and $\mu_\Xi$ as
\begin{equation}
\mu_\Lambda = \mu_s +\frac{1}{3} G 
\; \; \; {\rm and } \; \; \; 
\mu_{\Xi^-} = \frac{4}{3}\mu_s -\frac{1}{3}\mu_d 
-\frac{2}{3} G, 
\label{eq:mu}
\end{equation}
where $G\simeq$ 0.2 n.m. 
Thus the addition of the \J2hat contributions
gives a natural explanation of the inequality:
$ |\mu_\Lambda | < | \mu_{\Xi^-}| $.
\medskip
\begin{figure}
\begin{center}
\epsfig{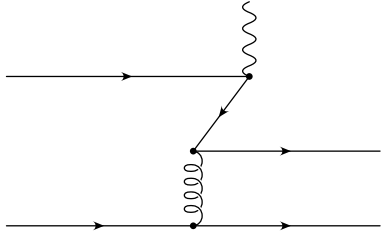}
\end{center}
\caption[]{Illustration of the effective
two-quark spin-spin correlation mediated by
an effective gluon exchange between the two quarks.
The top vertex is the photon quark vertex. }
\end{figure}

\medskip

The \J2hat contributions affect the magnetic moments 
of the other baryons as well. 
We show here that the \J2hat contribution is 
also helpful to resolve the ``$\Sigma$-$\Lambda$ problem",
a problem first pointed out by Lipkin\cite{lip99}
and further discussed by Thomas and Krein\cite{thomas00}.
Consider the magnetic moment ratio $R'$ defined by
\begin{equation}
R' \equiv
\frac{\mu_{\Sigma^+} + 2 \mu_{\Sigma^-} }
{\mu_\Lambda }
\label{eq:ratio1}
\end{equation}

The naive ``additive'' quark model gives $R' = -1$,
in glaring disagreement with the experimental value, 
$R'\simeq - 0.23$ \cite{lip99}.
Thomas and Krein have made a detailed study
of the Goldstone-boson (pion) contributions to $R'$,
using chiral perturbation theory 
as well as chiral quark models,
and emphasized the importance of
a proper treatment of the pionic contributions.
The leading non-analytic expression 
for the pion-loop corrections
was found to change $R'$ significantly.
%
%
We show here that there are additional important
contributions from \J2hat and that
these two effects combined
lead to a dramatic reduction of $|R'|$.
Here again we demonstrate our point
using the results obtained in the cloudy bag 
model \cite{hhfm88,kubodera}.
When the contributions from the Goldstone boson cloud
and \J2hat\ are included, $R'$ is given as
\begin{eqnarray}
R' =
&=&
\frac{\frac{4}{3} \mu_u -\frac{1}{3} \mu_s
+ \frac{1}{2} \delta \mu_\pi
+2\left(\frac{4}{3} \mu_d -\frac{1}{3} \mu_s
-\frac{1}{2} \delta \mu_\pi
-\frac{2}{3} G \right) }{\mu_s +\frac{1}{3} G } \nonumber \\
&\simeq & (-1) \left( 1 +
\frac{1}{2} \frac{\delta \mu_\pi}{\mu_s}
+\frac{4}{3} \frac{G}{\mu_s} \right)
(1 - \frac{G}{3 \mu_s} + \cdots )
\label{eq:ratio}
\end{eqnarray}
Here $\delta \mu_\pi = 0.59$ n.m. \cite{hhfm88} 
is the contribution 
from the pion-loop corrections and 
$G=0.2$ n.m.\ representing the \J2hat\ contribution
has already appeared in eq.(\ref{eq:mu}).
The value of $G/\mu_s$ 
consistent with eq.(\ref{eq:exp}) 
is $G/\mu_s \simeq -1/3$. 
Eq.(\ref{eq:ratio}) shows 
that both \J2hat and the pion-cloud contributions
substantially reduce the magnitude of $R'$,
and their combined effects 
essentially resolve the ``$\Sigma$-$\Lambda$ problem".

For the nucleon magnetic moments, 
the pion cloud contribution is an iso-vector contribution
and hence cannot explain the observed value,  
$\mu_n/\mu_p\sim -2/3$.
It has been demonstrated \cite{hhfm88,kubodera}
that the inclusion of the \J2hat contributions
leads to the observed value of $\mu_n/\mu_p$.

As mentioned, the effects of \J2hat\ 
can appear in weak-interation observables as well.
As an example, we discuss the ratio, 
\begin{eqnarray}
R''\equiv \frac{g_A/g_V (\Sigma^-\!\!\to\! n) }
{ g_A/g_V (\Lambda \!\to\! p)}  , 
\label{eq:GAGV}
\end{eqnarray}
where 
$g_A/g_V(\Sigma^-\!\!\to\! n)$ and 
$g_A/g_V (\Lambda\!\to\! p)$
are determined from measurements 
of the semileptonic decays of 
the $\Sigma^-$ and $\Lambda$ particles.
Experimentally, $R''= - 0.473 \pm 0.026$,
whereas the additive quark model gives
$R''=-1/3$.
Lipkin\cite{lip99} discussed this problem 
in combination with the above-mentioned ``$R'$ problem",
and stressed that it is impossible to simultaneously
explain the observed values of $R'$ and $R''$.
The argument in \cite{lip99}, however, is based 
on the essential assumption
that the $s$- to $u$-quark conversion 
can be described in the simplest ``additive" quark picture.
We therefore examine here
to what extent the \J2hat\  current changes 
$R''$ from the naive ``additive'' quark model value.
According to cloudy bag calculations 
\cite{hhfm88,kubodera},
the \J2hat\ current changes $R''$
as
\begin{eqnarray}
R'' = - \frac{1}{3} - 2 \frac{G^\prime}
{B^{\prime}} \simeq - 0.47 \; .
\end{eqnarray}
In the model used here, wherein
the quarks are relativistic
($m_u = m_d \simeq 10$ MeV), the ratio 
$g_A/g_V$ depends on the two integrals, 
$B^\prime$ and $G^\prime$,
involving the quark wave functions 
of the baryons; 
these wave function were calculated 
in Ref. \cite{hhfm90}.
The term with $G^\prime$ is 
the contribution of \J2hat \cite{hhfm88}.
It is found that this contribution 
changes $R''$ in the right direction
and can explain the experimental value of $R''$.
Thus, what was presented as a serious problem
in Ref.\cite{lip99} can be easily resolved
by including the chiral corrections 
due to the pion cloud {\it and} the correction 
due to the \J2hat current.

We have shown in this note the insufficiency
of the naive ``additive'' quark models. The 
\J2hat current presented here gives important contributions 
to both the baryon magnetic moments and the
hyperon beta decays.
It should be noted that
\J2hat also gives an important contribution to
the nucleon spin content \cite{hhfm95}
and leads to the result consistent 
with the Bjorken sum rule.
In conclusion, both the Goldstone-boson loop corrections
and \J2hat\   contributions play a significant role
in resolving the outstanding puzzles
regarding baryonic electroweak observables,
a feature indicating the importance of
quark-quark spin-spin correlations inside baryons.


\begin{thebibliography}{99}

\bibitem{thomas00} A.W. Thomas and G. Krein, Phys. Letters {\bf B481}, 21
 (2000); E.J. Hackett-Jones, D.B. Leinweber and A.W. Thomas, Phys. Letters
{\bf B489}, 143 (2000); D.B. Leinweber, D.H. Lu and A.W. Thomas, Phys.. Rev.
{\bf D 60}, 034014 (1999)

\bibitem{foot1} The 3-body term is expected to be negligibly small. 

\bibitem{foot2} A notable exception is Ref.\cite{morpurgo01},
which discussed the most general QCD-based  
parameterization of the baryon magnetic moments. 
See also Ref.\cite{af81}.  

\bibitem{morpurgo01} G. Dillon and G. Morpurgo, Phys. Rev. {\bf D53}, 3754
 (1996); G. Morpurgo, preprint hep-ph/0107049 (2001). 

\bibitem{af81} J. Arafune and M. Fukugita, Phys. Letters {\bf B102}, 437 (1981) 

\bibitem{hhfm88} H. H\o gaasen and F. Myhrer, Phys. Rev. {\bf D37}, 1950 (1988)

\bibitem{kubodera} K. Tsushima {\it et al.}
Nucl. Phys. {\bf A489}, 557 (1988);
Phys. Letters {\bf B205}, 128 (1988);
T. Yamaguchi {\it et al.}, Nucl. Phys. {\bf A500}, 429 (1989).

\bibitem{myhrer83} F. Myhrer, Phys. Letters {\bf B125}, 359 (1983)

\bibitem{cloet02} I.C. Cloet, D.B. Leinweber and A.W. Thomas,
preprint hep-ph/0203023 (2002)

\bibitem{lip99} H.J. Lipkin, preprint hep-ph/9911261 (1999)

\bibitem{hhfm90} H. H\o gaasen and F. Myhrer, Z. Phys. {\bf C48}, 295 (1990)

\bibitem{hhfm95} F. Myhrer and A.W. Thomas, Phys. Rev.
{\bf D38}, 1633 (1988);
H.H\o gaasen and F. Myhrer, Phys. Letters, {\bf B214}, 123 (1988);
 H.H\o gaasen and F. Myhrer, Z. Phys. {\bf C68}, 625 (1995).

\end{thebibliography}
\end{document}